# Experimental apparatus for non-contact resistivity measurements of the rock core plug based on magnetic induction

Pablo Diniz Batista[1§], Leduc Hermeto de Almeida Fauth[1§], Bernardo Coutinho Camilo dos Santos[2§], Willian Andrighetto Trevizan[2]§, Maury Duarte Correia[2§], Jorlandio Francisco Felix[3§*].

[1] Federal Institute of Education, Science and Tecnology of Brasília, QNN 26 Área Especial – Ceilândia, Brasília – DF, CEP: 72220-260 Brazil.

[2]Research Research Center of Petrobras (CENPES), Av. Horácio de Macedo, 950 Cidade Universitária, Rio de Janeiro, RJ CEP:21941-598.

[3]Universidade de Brasília, Instituto de Física, Núcleo de Física Aplicada, Brasília DF  70910-900, Brazil.

§ These authors have contributed equally to this work.
*Address correspondence to pablo.batista@ifb.edu.br

***Abstract*** **A new apparatus has been developed to measure the conductivity of rock samples. The probe, which consists of multi-coil transmitters and receivers doesn't require physical contact with the samples. The measurement is based on the induction principle. The measurement system is validated by using saline solutions and water-saturated sands of known conductivity. This work presents details of the development of a system of magnetic resistivity measurements by magnetic induction for petrophysical applications. The first application consists of measuring the resistivity of the core plug which is 0.038 m in diameter. Currently the system is operating properly at a frequency of 50 kHz with a current of up to 500 mA at 20 °C. During the study two types of samples were investigated: aqueous solutions with conductivities between 1 to 100 mS/cm and rocks. Several tests were carried out with the objective of investigating the performance of the instrument, such as the experiment to obtain sensitivity for the measurement system as a function of the current applied to the transmitter coil.**

## I.  INTRODUCTION

In general, the term petrophysics has been understood as being the study of physical and chemical rock properties and their interaction with fluids. In the context of the oil and gas industry it aims to optimize the process of extracting oil and its derivatives. For this reason, core analysis laboratories of rock samples obtained downhole provide essential data for exploration, evaluation and production of oil and gas reservoirs [1]. As core comes in a variety of lengths and diameters thus, the core analysis laboratories must be flexible enough to process the various types of core, such as whole cores and sidewall cores. Apart from these, there is a third type of rock sample known as core plug [2].

These plugs are taken as a representative subsample of the whole core and are useful in analyzing intervals between relatively homogeneous cores. A plug for analysis is typically 0.025 to 0.038 m diameter and 0.05 to 0.01 m long. Although core facilities typically accommodate a wide range of samples, core plugs are most frequently used for routine core analysis. There are sets of measurements which are normally carried out in laboratories, and the electrical characterization is one of the non-destructive geophysical methods commonly used [3].

This measurement can also be applied both in the field and in the laboratory. In the first case, a device commonly used is based on the magnetic induction for which the work principle is very simple: a time varying magnetic field produced by a coil induces eddy currents in the formation, and these currents produce a magnetic response that is measured by another coil on the tool.

On the other hand, in the laboratory, the electrical resistivity measurement has traditionally used the galvanic approach where electrical contacts are required. There are disadvantages which stem from the need for direct electrical contact between electrodes and core samples, as well as electrode polarization and contamination, for example [4].

Furthermore, and perhaps the most important motivation for this work is the recognition of the interest and demand by the oil and gas industry, to carry out resistivity measurements in wells and in the laboratory using the same physical principles. Thus, the development of a system of non-contact electrical resistivity measurements, to measure the electrical properties of core plugs based on magnetic induction, has great potential for use as an auxiliary tool in the drilling process for oil extraction.

Currently, as far as is known, there is no scientific instrument which carries out these measurements available on the market. However, over the last few years several proposals have emerged investigating the measurements of laboratory-scale resistivity which use the magnetic induction method. Even so, there are still few papers published on this research topic. One of these papers for instance, proposes the use of the electrical resistivity measurement to indirectly determine the porosity of the rock. A commercial measurement system composed of a single coil has been adapted to obtain resistivity data by induction. The output voltage is proportional to the quality factor of the coil tuned at a frequency of 2.5 MHz. The technique was used to determine the resistivity of the sample with a diameter of 0.09 m [5].

In another study, the soil resistivity is determined using a pair of transmitter and receiver coils operating at 100 kHz. The geometry of the measurement system is determined by an outer diameter of 0.10 m. The apparatus has been developed to measure the conductivity of samples at 100 kHz. The probe consists of coil-type transmitters and receivers and requires no physical contact with the sample. The sample is machined into a cylinder with an outer diameter of approximately 0.10 m and a 0.028 m inner diameter [6-8]. For unconsolidated samples such as soils or liquids, a sample holder of similar dimensions made of a non-conducting material such as Plexiglass can be used.

An other work proposes a non-contact resistivity system for whole core samples (with diameters between 1.75 and 5.25 inches). The apparatus uses only two coils, and in order to reduce the influence of the transmitter coil on the receiver coil the signal detected on the receiver coil must be processed before being correlated with the resistivity value of the sample. It was observed that one of the issues is to try to reduce the size of the coil diameter, and therefore the size of the sample. That is, reducing the size of the coils implies the decreasing of amplitude of the signal detected on the receiving coil [9]. Finally, a more recent work presents an interesting system of non-contact resistivity measurements with applications in geophysics operating at a frequency of 51.28 kHz. The apparatus is cylindrical with 0.61 m long by 0.10 m in diameter. The differential of this instrument is its capacity to investigate samples with anisotropy. At present work the measurement system is standardized to 0.10 m long and 0.038 m in diameter to meeting the standard established by the Petrobras research laboratory [10,11]. Additionally, for system used here was developed a RF amplifier to supply the specific AC current at 50 kHz, which is applied to the transmitter coil.

## II. Mathematical model

The operating principle of the contactless measurement system is based on the induction theory and the mathematical model for the measurement of electrical conductivity by electromagnetic induction, taking into account a system of measurements basically composed of two coils [12].

There are many ways to sense magnetic fields, most of them based on the intimate connection between magnetic and electric phenomena. In this apparatus, the measurement sensors are inductive coils in which the principle is described by Faraday's law of induction. If the magnetic flux through a coiled conductor changes, a voltage proportional to the rate of change of the flux is generated between its leads. Induction coil sensors are one of the oldest and most well-known types of magnetic sensors. It is practically the only one that can be manufactured directly by a user because the method of coil manufacture is simple and the materials are easily available [13,14]. The response of the coil-type probe can be accurately predicted when the conductivity and the dimension of the samples are known [7]. The current in the transmitter coil creates eddy currents in the object at a known frequency. Then, the eddy current generated in the object induces a voltage in the receiving coil. From this configuration, the proposal of the measurement system basically consists in monitoring the voltage generated in the receiving coil as a function of the resistivity, thus obtaining a conductivity voltage calibration curve. The induction technique applied to the electric characterization is based on the first work presented by Moran e Kunz [12]. According to his theory, the induced voltage in the receiver coil is given by

$$V = \frac{TR(\pi a^2)^2}{2\pi} Ii\omega\mu(1 - ikL)\frac{e^{ikL}}{L^3} \qquad (1)$$

where R and T are the number of turns wound as the receiver and transmitter on a core of radius $a$, respectively. $\omega$ is the frequency and $\mu$ is the magnetic dipole moment. Separating the real and imaginary part of $V$, expanding Eq. (1) into real and imaginary components as power of $KL$,

$$-V_R = K\sigma\left(1 - \frac{2}{3}\frac{L}{\delta} + \cdots\right) \tag{2}$$

$$V_X = K\sigma\frac{\delta^2}{L^2}\left(1 - \frac{2}{3}\frac{L^3}{\delta^3} + \cdots\right) \tag{3}$$

$$K = \frac{(\omega\mu)^2(\pi a^2)^2}{4\pi}\frac{TR}{L}I$$

and

$$\delta = \sqrt{\frac{2}{\omega\mu\sigma}} \tag{4}$$

The term $\delta$ is the depth of penetration, which provides the magnitude of the penetration depth of the magnetic field in the conductor. Understanding the theoretical model for the contactless resistivity measurement is fundamental to optimize the quantities involved in the experimental measurement system, obtaining a linear relationship between the voltage at the receiving coil and the resistivity of the sample. Finally, from the voltage response we can study samples with unknown resistivity.

From these first considerations it is important to highlight that the geometric factor has a dependence with the square of the frequency, therefore, the smaller the frequency of the signal the less the sensitivity of the measurement system. Another important point is to recognize that the real voltage response is linearly related to the electrical resistivity; however, the imaginary voltage response is independent of conductivity and therefore represents only the direct mutual coupling between the transmitter and receiver coils regardless of their surroundings [10,11].

### III. EXPERIMENTAL APPARATUS

**Figure 1** shows schematic diagrams of the proposed apparatus developed to perform the experimental measurements in order to determine the resistivity of rock plugs with 0.0381 m in diameter using magnetic induction. In general, the experimental system can be divided into two parts. The first part is related to the electronic instruments used: (a) a signal generator; (b) a lock-in amplifier and (c) a digital multimeter. For now, all these instruments are commercial; in particular, the MFLI Lock-in Amplifier is the most import instrument in this setup. However, we have developed, designed and manufactured a radio frequency (rf) power amplifier. The second part comprises a 0.25 m long cylindrical nonconductive sleeve sized to fit a 0.038 m diameter geological core or holder with brine, which is also used to fix the coils that make up the mechanical part of the measurement system; two coils are used as a transmitter (20 turns each), one receiver (20 turns), a buck (10 turns) and one as a reference signal (25 turns). The buck coil is designed to mitigate the coupling between the transmitter and receiver coils by minimizing the voltage generated on the receiving coil **[10].**

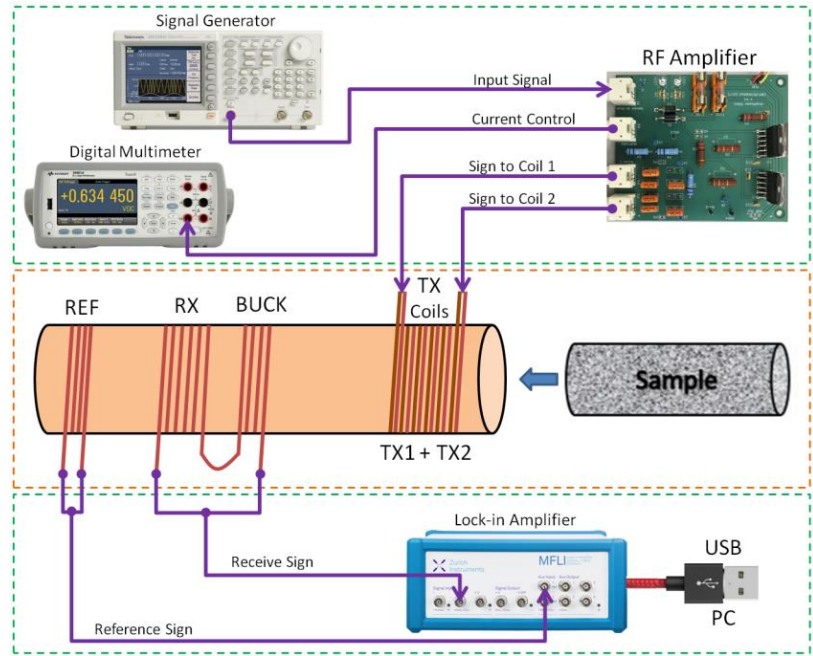

**Fig. 1.** A schematic of the experimental apparatus for non-contact resistivity measurements based on the magnetic induction principle. Basically, the system is composed of a signal generator, which is used to generate an input signal in the RF amplifier, which, in turn, amplifies the AC current signal. This signal is then applied to the transmitter coils (TX1 and TX2). Finally, a voltage signal is generated on the receiving coil which is measured by lock-in amplifier.

In general, a sinusoidal signal with a frequency of 50 kHz from the signal generator is amplified by a power amplifier and the output provides the electrical power to two transmitter coils. This allows the primary magnetic field produced by the transmitter coil to induce an eddy current in the sample. This current, in turn, generates a secondary magnetic field, which is inversely proportional to the resistivity of the sample. In addition, it is important to recognize that the receiver coils detect the magnetic fields produced by both the transmitting coils and the induced current in the sample. In addition, the voltage produced by eddy currents is usually much smaller than the primary one. This way, the measurement of this secondary field is not so trivial, it requires careful mechanical adjustment of the buck and receiver coils **[10-11].** In order to measure the secondary field accurately we found it necessary to remove the primary field and amplify the difference to measurable levels. The design of the receiver coil is crucial for the overall performance of the system since digital processing requires an impractically high dynamic range of the analog-to-digital converter (ADC). Therefore, it is highly recommendable to reduce the induced voltage in the unloaded system while preserving maximum sensitivity to conductivity changes. Among the various configurations, it is possible, for example, for the subtraction of the signals in a pair of differential coils (gradiometer) as proposed in different works **[15,16].** After this the voltage induced in the receiver coils, it can be measured using a lock-in amplifier. The reference signal for the lock-in amplifier comes from the reference coil. By comparing the phases between the receiver signal and the reference signal, the lock-in amplifier measures the amplitude of the in-phase and the out-of-phase components of the received signal. In the future, the main idea is that all the commercial equipment used in this work be integrated into a single scientific equipment. In this case, it will be necessary to develop a Lock-in Amplifier, a low frequency signal generator and an AC current meter **[17, 18].**

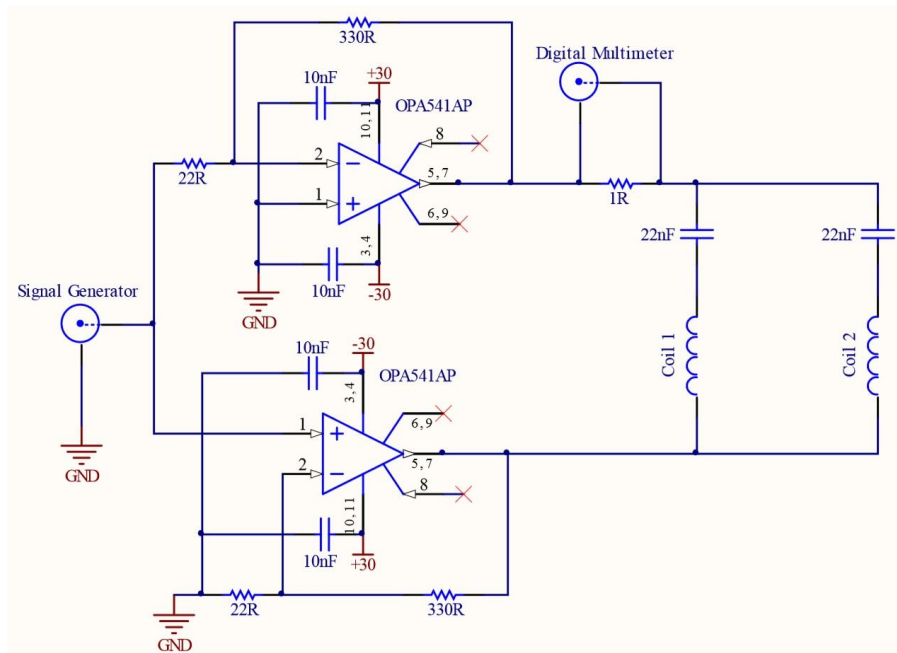

**Fig.2.** Electrical scheme of the RF power tuned at 50 kHz for a AC current of up to 1A. The Circuit Board produced using this scheme is shown in the upper right-hand corner of Figure 1.

**Figure 2** shows the electrical schematic for the first prototype referring to the RF power amplifier used in the measurement system. It is important to point out that we designed this circuit board (also shown in the upper right-hand corner of Figure 1) to meet the specific demand of our experiment. Therefore, the power amplifier was designed to provide enough AC current in the transmitter to generate the time varying magnetic field that will create the eddy currents in the sample. In this case, the main idea is to use two operational amplifiers configured in a bridge to provide currents up to 1 A, because in this configuration the maximum voltage applied to the load can be up to twice the supply voltage. Both operational amplifiers shall be mounted in both inverting and non-inverting configurations, providing that the voltage at the load is equal to twice the supply voltage. This is the first trick used to increase the power load and therefore the magnetic field strength in the sample. Furthermore, to increase the current and, for instance, the sensitivity at low frequency we placed a capacitors bank in the series with the excitation coil to produce a resonance at 50 kHz. The second trick is to use two RLC circuits in parallel, because the power dissipated by the coil will be divided in two, thus reducing any noise during the measurements. Since the first version this electronic circuit has been designed from the OPA541 operational amplifier capable of operating with voltage sources of ± 40 V and currents up to 5A. With the electronic schematic and by using the program Altium a first version of the printed circuit board could be elaborated. In this printed circuit board the positioning of the components is adjusted so that heat dissipation in the electronic circuit as a whole is efficient.

A user friendly interface was developed using C++ exclusively for conductivity measurements with a focus on daily use in a research laboratory. The developed program uses a library provided by the manufacturer of the MFLI Lock-in Amplifier for communication through the USB port. The ziAPI library for C language is a simple and robust interface that can be executed on most platforms allowing configuration and retrieval of parameters as well as data. Since C is a low-level programming language, the development cycle is slower when compared to other programming environments. However, the result is very gratifying.

The differential of this product is that it makes resistivity measurements a simple and fast procedure. For this the procedures should not require the user details of the operation of the scientific instrument. First, the program configures all MFLI Lock-in Amplifier parameters without any user intervention, for example, gain, integration time, phase value, and several other parameters are selected. Then the real and imaginary voltage is highlighted by the user. From these measurements the program allows for the carrying out of calibration using two samples of known conductivities. In this case they are 1 mS/cm and 100 mS/cm. After this procedure the program determines directly through the actual voltage the conductivity value measured by the experimental apparatus presenting the result both visually in a colour bar and in mS/cm. In addition, the program allows the values of real and imaginary voltage as well as measured conductivity to be saved in a text file for processing.

## IV. PREPARATION OF THE SAMPLE

The brine solutions used in the experiments have conductivity between 1 and 100 mS/cm, whose values correspond to those found in oceans. In addition, to show the robustness of the measurement system, the calibration curve is obtained using eleven solutions with 10 mS/cm intervals. All solutions with known conductivity solutions are prepared using the commercial DIGIMED conductivity meter in order to investigate the measurement system. Also, two conductivity cells are required for this experiment: one for conducting low-conductivity (DMC-010) and one for high (DCM-100) measurements. The accuracy is ±1% FS. A sample holder was designed to measure a uniform liquid sample in order to verify the accuracy of the system by measuring uniform saline solutions with known salinities. In this case, the sample holder with a diameter of 0.380 m will be used to store an aqueous solution so that it can be inserted into the measurement system. This holder can also be used to measure conductivities of soils, sands, and other unconsolidated materials. The system used for sample saturation is basically composed of a vacuum pump and a desiccator.

## V. Results

First shown is the frequency response of the transmitter coils connected to the RF power amplifier developed to operate close to 50 kHz. **Figure 3-a** shows the electric current measured in a shunt resistor ($R_2 = 1$ Ohm) using an Agilent Digital Multimeter as a function of the frequency of the signal applied to the input of the RF power amplifier. Note that signals with frequencies between 10 kHz and 150 kHz were used for this experiment and the results show that the resonant frequency of the coupled coils is at 70 kHz. The resonance frequency is shifted to the right adjusting the capacitor values in such a way that the current (approximately 500 mA) required for the measurement system is able to detect aqueous solutions with conductivity between 1 mS/cm and 100 mS/cm. In addition, the results demonstrate in detail the proper stability of the power amplifier for different current values at 50 kHz (see **Figure 3-b**).

During tests with the measurement system it was observed that the signal at the output of the power amplifier is out of phase with respect to the input signal of the signal generator due to the intrinsic characteristics of the operational amplifiers used, as shown in **Figure 3-c**. **Figure 3-d** illustrates the delay time increases linearly with the signal frequency obtained using the TDS2400 oscilloscope. This effect must be taken into account during the calibration of the measurement system, therefore, as a first option we recommend the use of a reference coil to provide a voltage to the Lock-in Amplifier as a reference signal. Unfortunately, the phase adjustment has to be performed manually, that is, the position between the buck and receiver coils relative to the transmitter coil, as well as the number of turns, should be optimized so that the measured voltage is minimized in the absence of a sample. However, depending on the presence of parasitic capacitance, due to the cables for instance, this procedure is not sufficient to adjust the measurement system. Therefore, this is the most delicate part of the experimental system which must be carried out cautiously and patiently in order to figure out a configuration that maximizes the response of the measurement system [7].

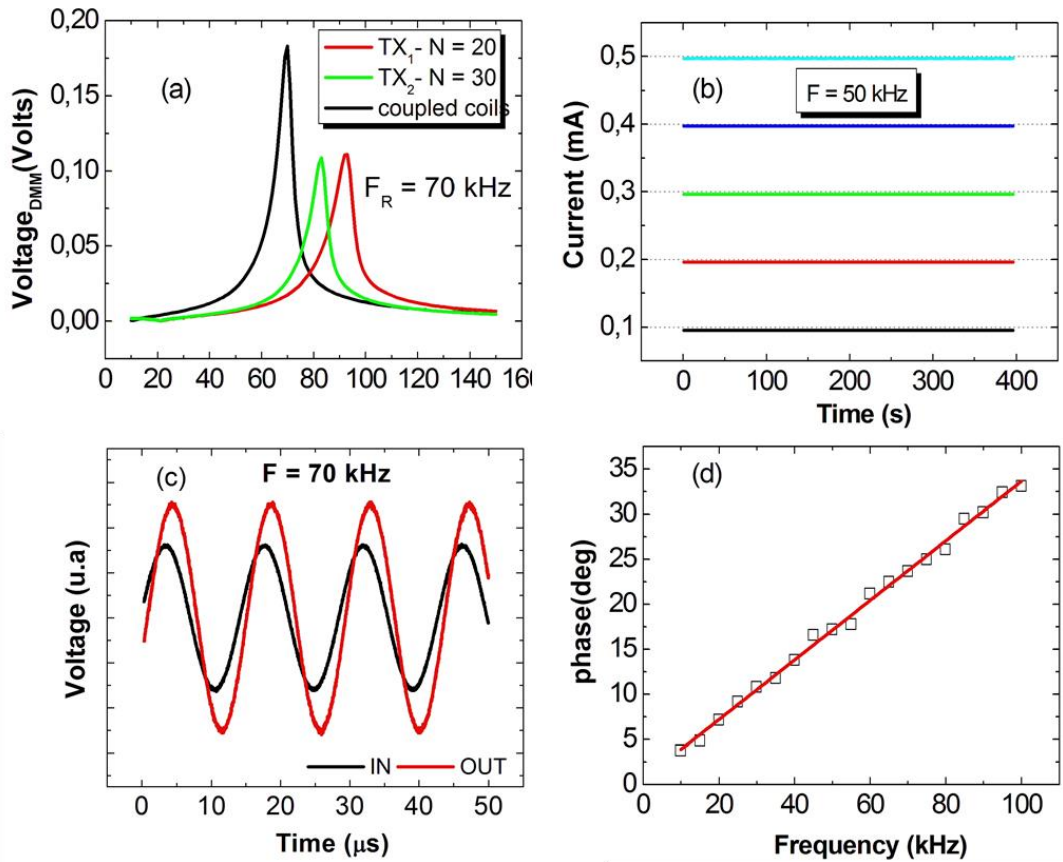

**Fig. 3.** (a) Frequency response of the transmitter coils. (b) Load current as function of time for different power values. (c) Input voltage delay with respect to the output on the RF power amplifier. (d) Delay time as function of frequency of the signal generator. Note that the signals are on different scales due to the gain of the amplifier

**Figure 4** shows the response of the measurement system for saline solutions with conductivity ranging from 1 mS/cm to 100 mS/cm at intervals of 10 mS/cm. The system operates at a frequency of 50 kHz, the gain in the Lock-in amplifier is set to 200 and the integration time in 200 ms with an order 2 RC filter. The monitored real voltage depends on the resistivity of the solution immersed in the sample holder. In order to investigate the effect of the electric current on the response of the measurement system this same experiment is repeated by adjusting the current to 100, 200, 300, 400 and 500 mA (**Figure 4-a** to **Figure 4-e**). In addition, the value of the electric current is automatically adjusted by the program by changing the voltage value applied to the RF amplifier through the signal generator. Furthermore, the imaginary voltage (not show here) does not have a clear pattern as a function of this magnitude.

To determine the calibration curve or response voltage, the developed program first considers only the moment when the sample is present in the measurement system, thus disregarding the voltage value during sample exchange. Then, the program calculates the average of the real voltage for each conductivity value. The results show that the response is directly proportional to the electrical conductivity of fluid inside the sample holder, confirming that the measurement system is performing as expected. These data also exhibit a straight line relationship between the measurement of the secondary field and the fluid conductivity and can be used to calibrate the output in terms of the resistivity of saline solutions (see **Figure 4-f**).

The results show that the values for the sensitivity of the measurement system are 14.54, 26.60, 40.83, 52.97 and 63.39 µV/(mS/cm) respectively. As expected, the sensitivity is directly proportional to the electric current applied to the transmitter coils. For example, when a current of 500 mA is applied to the transmitter coils the voltage increases by only 63.39 µV for every 1mS/cm, thus indicating the importance of a stable measurement system and a Lock-in amplifier with a resolution of at least the order of µV.

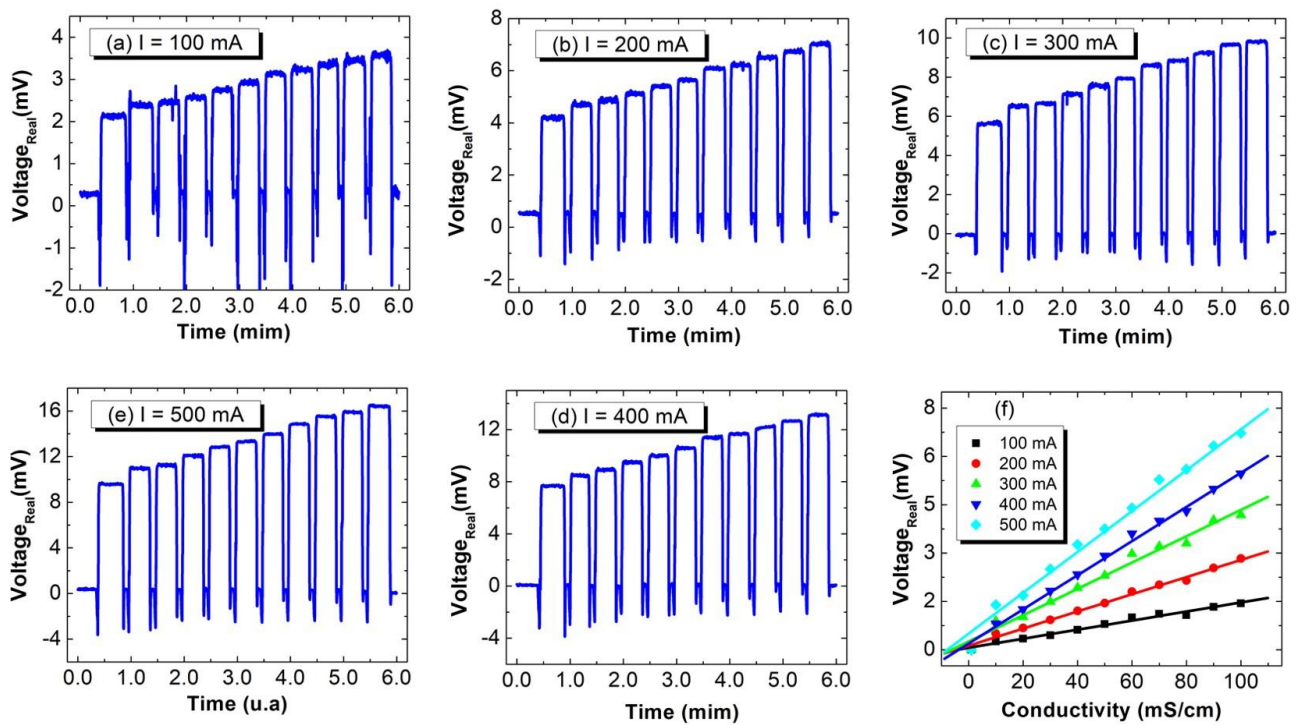

**Fig. 4.** Measurement of real voltage for samples with conductivities from 1mS/cm to 100 mS/cm. The current applied to the transmitter coil has the value of (a) 100, (b) 200, (c) 300, (d) 400 and (e) 500 mA and (f) calibration curves as function of current.

Thus, one of the ways to increase the sensitivity of this measurement system, if necessary, consists of adding scientific and technological efforts to increase the operating power of the RF amplifier for this operation frequency. On the other hand, it is more elegant in terms of scientific research, to try to use other magnetic sensors which are more sensitive than the inductor coils, such as atomic magnetometers for example [19].

Finally, the methodology proposed in this work is used to determine the resistivity of a rock taken from one of the wells drilled by Petrobras. First, the rock is saturated for 30 minutes in a vacuum in a saline solution with conductivity in the order of 100 mS/cm.

Thereafter, the experimental apparatus can be used to measure the voltage generated on the receiving coil due to the saturated rock. In this case, the real voltage from the rock is compared with the real voltage obtained using two standard solutions of 1 and 100 mS/cm. The results obtained from this experiment and presented in **Figure 5**, indicate a conductivity in the order of 52.74 mS/cm (or 0.5274 mho/cm) for the rock investigated in this work.

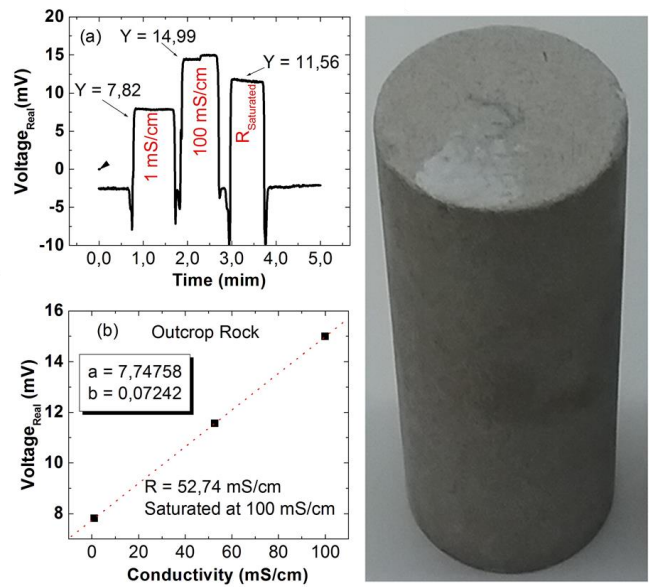

**Fig. 5.** Measurement of the resistivity of the cylindrical rock core sample (0.1016 m long x 0.0381 m diameter) shown on the right side. For this experiment the rock is saturated in a saline solution of 100 mS/cm for 30 minutes.

## VI. CONCLUSION

This work presented in detail the technological and development of the first prototype of a laboratory apparatus for the measurement of conductivity in 0.38 m diameter rock plugs using the magnetic induction method. For a current of 500 mA and a gain of 200 the sensitivity of the experimental apparatus is in the order of 64 μV/(mS/cm). Several experiments were carried out in order to verify the robustness of the measurements. The results demonstrate the fundamental importance of (i) the decoupling of the magnetic coils, and (ii) the phase adjustments. It is recommended that this first prototype be used to perform tests on a larger scale when considering different classes of rocks. In addition, for the consolidation of this instrument, these measurements of conductivity by electromagnetic induction should be corroborated with the contact technique. This allows for the definition of the differences between the two approaches. As a future perspective, in terms of scientific instrumentation, this experimental system will be expanded to operate with frequencies up to 500 kHz. It must be emphasized that currently there is no commercial equipment or any scientific work that meets all the characteristics discussed in this work. Finally, it is important to consider the development of part of the scientific instrumentation used to assemble this experimental apparatus, since many works use commercial equipment to perform non-contact conductivity measurements. Furthermore, because the measuring probe is not in contact with the sample, this system is especially desirable for measuring hazardous substances [7], and it can also be used to measure contaminated soils.

**ACKNOWLEDGMENT**

Thanks to all who contributed to this work, they are: Aridio Schiappacassa, Edgar Monteiro da Silva, Barbara Aguiar and José Eduardo. This work is the result of a cooperation between CENPES and CBPF (Nº 0050.0093372.14.9; PT-160.09.11970; SAP-4600479282).

**Pablo D. Batista** receives his PhD degree in Physics Applied to Medicine and Biology from the University of São Paulo in 2009. During the doctorate a scholarship from the DAAD for doctorate sandwich at the Paul Drude Institut (Berlin-Germany). He is currently a professor at Federal Institute, Science and Tecnolog of Brasília and has experience in the field of Physics, with emphasis on General Instrumentation Specification in Physics, acting in our main subjects: scientific instrumentation, microcontrollers, C/C++ language, pH sensors, EGFET, lock-in amplifier, optical sensors for glucose detection.

**Leduc H. de A. Fauth** receives his a degree in Physics from the Fluminense Federal University (2009) and a Master's degree in Physics with an emphasis on scientific instrumentation by the CBPF (2017). He has been working at the Navy Research Institute since 2012, being part of the Inertial Systems Division and working on auxiliary sensors.

**Jorlandio F. Felix** Jorlandio Francisco Felix received the Ph.D. degree from the Federal University of Pernambuco, Recife, Brazil, in 2012. He went to The University of Nottingham, Nottingham, U.K., in 2011, where he was involved in one deep level transient spectroscopy techniques setup. He has been involved in semiconductor materials and devices for many years. His current research interests include inorganic/organic semiconductors and their applications to solar cells and photonics. In the field of applied physics is working in the development of techniques for electrical characterization for determining rock properties of core and surface samples.

**Maury D. Correia** received his PhD degree in Physics from Brazilian Center for Research in Physics (CBPF), Brazil in 2015. He work at Research Center of Petrobras (CENPES) since 2010, his main subject research area are physical statistical models to petrophysical and geological properties of porous media like magnetic relaxation and electrical response.

**Bernardo C. C. dos Santos** received his PhD degree in Physics in 2009. He work at Research Center of Petrobras (CENPES) since 2010 and the head of CENPES NMR laboratory since 2011. His main research subjects are NMR petrophysics, statistical mechanics based models for porous media and its electrical properties.

**Willian A. Trevizan** received his PhD degree in physics from USP, where he develops techniques and models of Nuclear Magnetic Resonance (NMR) petrophysical petrography and  characterization of reservoir rocks of interest in the Oil and Gas Industry. He is the coordinator of the Research Laboratory of the Petrobras Research Center (CENPES) and has long experience in inversion in magnetic relaxation data on porous media, NMR acquisition protocols under of reservoir (pressure and elevated temperature) besides acting in the fundamental agreement and geoscientific of rocks of great economic interest such as the pre-salt rocks and turbidity sandstones.